# Drawing Out Legal Risks: Co-Designing with Lawyers to Predict and Manage Legal Uncertainties of Medical AI Tools

**Gennie Mansi**
Georgia Institute of Technology,
School of Interactive Computing

**Julia Kim**
Georgia Institute of Technology,
School of Interactive Computing

**Michael Rosenbloom**
Harvard School of Law,
Cyberlaw Clinic

**Mark Riedl**
Georgia Institute of Technology,
School of Interactive Computing

## ABSTRACT

While there's optimism around medical AI tools due to their abilities to adapt from user-to-user and across environments, these new abilities complicate how people and organizations are able to predict and manage risk based on existing laws and regulations. Lawyers are trained to identify potential legal outcomes, but they lack technical AI knowledge, making it difficult to translate their expertise to creators and users of AI tools. We contribute insights from our co-design process with U.S. lawyers to identify and translate ways to predict and manage risks of medical AI tools. We present the visualizations we developed through two years of cross-disciplinary efforts and thereby illustrate our findings about how legal risks are determined and our strategies for people and organizations to predict and manage these risks. We offer insights about leveraging lawyers' expertise to understand, predict, and manage legal risks.

## INTRODUCTION

In the United States, it is critical for people and organizations who create or use medical technology to **predict and manage legal risks**. Legal risks depend on interrelated, unpredictable, and complicated factors. For creators—such as developers or hospitals—and users—such as doctors—many factors impact legal risk, including the state that they're working in, who is hurt, intended use, and standard operating procedures [7]. It is also challenging to predict the ultimate costs, such as financial and reputational consequences [17, 24].

Creators and users analyze potential risks based on current regulations, and they proactively strategize to minimize negative impacts before, during, and after medical technology is made. For example, creators may implement a well-known, structured risk management program and ongoing training for current employees [11]. Users, such as doctors, may manage their risks by attending all required trainings for technology, reporting errors according to hospital policy, and ensuring their use of a technology remains in line with broader professional standards for care [4, 23].

Despite the procedures for minimizing legal risks of new medical technologies, people and organizations are **struggling to connect AI tools' abilities to laws and regulations** with which they must comply. The ability of AI-based medical tools to adapt from user-to-user and across different environments is a significant strength, but marks a stark shift from prior medical technologies whose performance and variance could be fully tested and predetermined [18, 20]. Widely respected medical groups warn doctors against the use of medical AI tools because it is unclear how the tools' new abilities connect to their existing legal responsibilities [21, 25]. Many hospitals are wary of medical AI tools because it is unclear how they may share responsibility when doctors' uses of these tools result in harms to patients [13, 18]. Finally, legal scholars warn about gaps between AI tools' abilities and existing regulations, which may leave patients, their families, and caregivers without pathways for recourse from harms resulting from their use [20, 26].

**Lawyers may be able to** help people and organizations **identify potential legal risks** of existing medical AI tools or situations, but they are **not trained to articulate design requirements** for medical AI tools. Prior work has highlighted the value of engaging with lawyers in co-design methods to incorporate their expertise into the design of AI tools for high-stakes environments [12, 16]. However, lawyers differ from technologists in their understanding of what AI tools can do [8] and are not trained to articulate software features to meet users' needs. As a result, it is challenging to translate lawyers' expertise into actionable steps for predicting and managing legal risks throughout the creation and use of medical AI tools.

In this pictorial, we contribute insights from our assets-based, co-design process to leverage the expertise of U.S. lawyers. We present the visualizations—flowcharts, diagrams, and time lines—we co-created with lawyers to construct a shared understanding of the information needed to analyze legal risks and to ideate strategies for predicting and mitigating risks. From our co-design process, we observe that legally-relevant information depends on how laws and regulations recognize harm and on the dynamics between the people and organizations creating and using these tools. We find ways for people and organizations to predict and manage their legal risks by documenting legally-relevant information where problems are likely to occur and taking action throughout the creation and use of medical AI tools. Based on our findings, we discuss how creators and users of medical AI tools can leverage lawyers' expertise.

## LAWYERS' SKILLS IN RISK ANALYSIS

Lawyers develop persuasive arguments that connect abstract legal theories or principles to specific legal situations. As a part of their job, lawyers regularly engage in risk analysis—identifying the legally significant information in a specific situation and how it can lead to potential legal outcomes. For example, before proceeding with a particular action, a client may want to know the full range of legal risk they may face in a particular circumstance.

Lawyers are trained to avoid speaking 'off-the-cuff' about legal risks. Instead, they write legal memorandums, or ***legal memos***, standalone documents that identify potential and likely claims as well as information needed for further investigation [22]. Legal memos analyze the entire situation a client is in, presenting the full range of legal possibilities resulting from the specific facts of a situation [6].

To write legal memos, lawyers reference ***fact patterns***, narrative descriptions of the relevant facts in a legal case. They analyze the fact pattern to understand how legal structures and case law may apply to the facts of their client's situation [6]. They can then identify potential legal outcomes and costs, such as financial and reputational costs, from a legal situation. In our co-design process, we leveraged lawyers' skills in risk analysis to connect the design of medical AI tools to current laws and regulations.

## USING VISUALIZATIONS WITH LAWYERS

**Visualizations can support cross-disciplinary communication** needed to translate lawyers' expertise into actionable steps for those without legal expertise. There are several challenges for engaging in cross-disciplinary co-design with lawyers. Lawyers may not understand AI systems' capabilities [8]. There are also noted terminological differences among fields, such as computer science and law [3].Design approaches with lawyers have successfully used visualizations to support cross-disciplinary conversations [9, 10, 19]. For example, the Wikimedia foundation used design approaches to re-create their trademark policy and support discussion among its lawyers, contributors, and users around the new policy [9]. While this work demonstrates the potential for using design approaches with lawyers to innovate legal changes with new technology, there is still need for work to explore ways of creating specific recommendations for the design of AI systems to predict and manage legal risks.

Note: The content of this paper does not constitute legal advice and should not be construed as a substitute for any legal services from an attorney.

## TERMINOLOGY

This paper preserves the technical language that lawyers use. This sidebar provides definitions for the legal terms referenced throughout the paper.

- **Entity**: a legally recognized organization or structure that can enter into contracts, sue or be sued, and hold property in its own name, separate from the individuals who own or work for it; examples include corporations and non-profits
- **Party**: an individual, organization, or entity involved in a legal situation or proceeding
- **Claim**: an assertion by one party against another because of the violation of a legal right
- **Harm**: injury or damage suffered by one party
- **Plaintiff**: the person or organization that starts a lawsuit because they believe they've been wronged and want the court to help fix the problem or award them money
- **Defendant**: the person or organization accused of wrongdoing in a lawsuit and who defends themselves against the claims by the plaintiff
- **Facts**: the circumstances, events, actions, and conditions that the parties state have occurred
- **Liability**: the legal responsibility for something that went wrong
- **Due Diligence**: the reasonable investigation, care, and review that a person or organization can be expected to undertake
- **Legal Exposure**: the potential risk or liability that a party faces in litigation, including the potential damages, penalties, or adverse outcomes
- **'Recognize'**: the formal acknowledgment of a right, claim, relationship, or legal principle
- **Personal Health Information (PHI)**: information about health status, healthcare services, or healthcare payments that are linked to a person

# METHODS

Using an asset-based [27] approach, we worked closely with two lawyers (who we refer to as **project partners**)—one in tort law and medical technology, the other in policy and platform regulation—to co-create interview protocols eliciting lawyers' skills in risk analysis. We then conducted one-on-one interviews with 8 other lawyers (who we refer to as **participants**) to identify information impacting legal risks from the use of medical AI tools for seven unique fact patterns matching their different areas of expertise in health law (see 'One-on-One Interview Sessions' under *Co-Design Process* below).

With our project partners, we analyzed interview data and co-created visualizations—flowcharts, diagrams, and time lines—to construct a shared understanding of the interview data and ideate strategies. From each interview, we co-created a flowchart of the information needed to analyze risk and a diagram of the dynamics between parties. Together, we connected the flowchart to the creation and use of medical AI tools. Using these visualizations, we co-created a time line of strategies for predicting and managing legal risks.

## CO-DESIGN PROCESS

We present each step of our co-design process below and note which findings and visualizations relate to each step.

**Co-Create Interview Protocols**

With our two project partners, we co-create fact patterns (example to the right) and interview protocols to leverage lawyers' assets.

**One-on-One Interview Sessions**

We conduct one-on-one interviews with lawyers. Each lawyer receives a unique fact pattern matching their expertise, except for two participants who overlapped in their area of focus. We collect and anonymize legal memos; interviews are recorded and transcribed.

**Understanding the Information Lawyers Need for Risk Analysis**

We analyze the legal memos and interview transcripts, co-creating a flowchart and diagram for each fact pattern.

**Findings 1-2 and their visualizations are from this step.**

**Connecting Lawyers' Risk Analysis to the Creation and Use of Medical AI Tools**

We annotate the flowchart to identify connections with the design of medical AI tools.

**Finding 3 and its visualizations are from this step.**

**Co-Creating Strategies to Predict and Manage Risk**

Based on the annotated flowchart and diagram, we co-create strategies for predicting and managing risk.

**Findings 4-6 and their visualizations are from this step.**

## INTERVIEW DESIGN

To engage participants' risk analysis skills, we asked them to write legal memos based on fact patterns involving doctors' uses of medical AI tools. Because lawyers are trained to avoid spontaneous responses, we emailed participants a fact pattern and asked them to respond asynchronously. We conducted 15-minute follow-up interviews within 1-2 weeks to discuss their memos. **Below we annotate the fact pattern for our running example.**

Each fact pattern incorporated 4 elements

**Clinical opportunities for medical AI tools**

We use opportunities for AI identified as clinically useful by the National Academy of Medicine (NAM), a leading US health policy advisory group [14]

**Relevant areas of health law**

We include information tied to a specific relevant area of health law

**Clinical concerns for medical AI tools**

We include concerns for medical AI applications that the NAM highlighted

**Specific prompt to consider risks for different parties**

We include explicit prompts asking for specific legal information from different parties

A hospital's electronic health record (EHR) system has an AI-based translation tool to help support communication between doctors and patients by translating doctors' medical notes, patient education materials, and messages into a patient's preferred language. The doctor selects which documents to translate and send to patients.

The translation tool works by scrubbing personally identifiable and protected health information from the document before sending it to an external large language model (LLM) to be translated. The LLM is from a major, reputable company (that is not the EHR company). After the translated content is returned, the tool reinserts scrubbed data back into the document before sending it to the patient. The EHR company claims the tool complies with patient privacy laws and regulations because it scrubs any sensitive patient information before sending it to the external LLM.

A physician uses this tool to translate a message into a language that they do not speak or understand. The text they ask the tool to translate contains protected health information. Later, it is discovered that protected health information was not being properly scrubbed and was actually sent to the external LLM.

Assume that the patient comes to you for legal assistance.

1. What claims are there and against who? Are there facts that are not given in the description that might determine the outcome? If so, what are those facts and what are the range of outcomes?

Now assume you're the physician's lawyer.

2. How would you defend any claims made against the physician? Are there facts that are not given in the description that might determine the outcome? If so, what are those facts and what are the range of outcomes?
3. What would you have advised the physician to do (or not) beforehand to minimize the risk of being sued beforehand?

This fact pattern focuses on two opportunities: routine patient communication and translating educational materials and health information [14]

We specify an area of health law corresponding to the clinical concerns

We include data privacy and security as a major risk for the clinical use of medical AI tools

We ask the participant to provide information for the patient and doctor

# GUIDE TO THE PICTORIAL

This pictorial presents visualizations we created in our study, analysis from those visualizations, and findings. The visualizations that we co-designed represent the many factors impacting legal risk, as translated from participants' memos. In this pictorial, we annotate the visualizations with our analysis and the themes we surfaced around predicting and managing legal risks. The annotations reveal rich details of lawyers' tacit knowledge and the challenges of predicting and mitigating legal risks.

The use of visualizations in our work and this pictorial is motivated by the need for shared accessibility and explorations among cross-disciplinary experts. Prior work has called for the creation of design artifacts that can serve as boundary objects, promoting shared understandings among the legal and technology communities [16]. The pictorial describes what we learned from using the visualizations as boundary objects with our project partners, and communicates our findings with the broader research community. We overlay hand drawn illustrations created by the first author from her reflections on emergent themes surfaced through the co-design process. **Layering the annotated study artifacts and hand drawn illustrations, we literally draw out the legal risks, informing a richer understanding of the legal complexity embedded in the design of medical AI tools and advancing practices for collaboration.**

Every Finding has 3 visual components: a study artifact, annotations of our analysis, and illustrations. Below we describe and provide examples of these three components and unpack the visual elements used to present our Findings.

### Study Artifact

The visualizations we co-created with study partners; made of non-hand drawn digital shapes and outlined with black boxes

### Annotations

Hand drawn lettering, bubbles, arrows, and connectors contextualizing salient characteristics of the artifacts with our analysis; on this page, we provide examples of annotations in shades of green and teal

### Illustrations

Drawings by the first author reflect emergent themes surfaced through the co-design process; not on every page

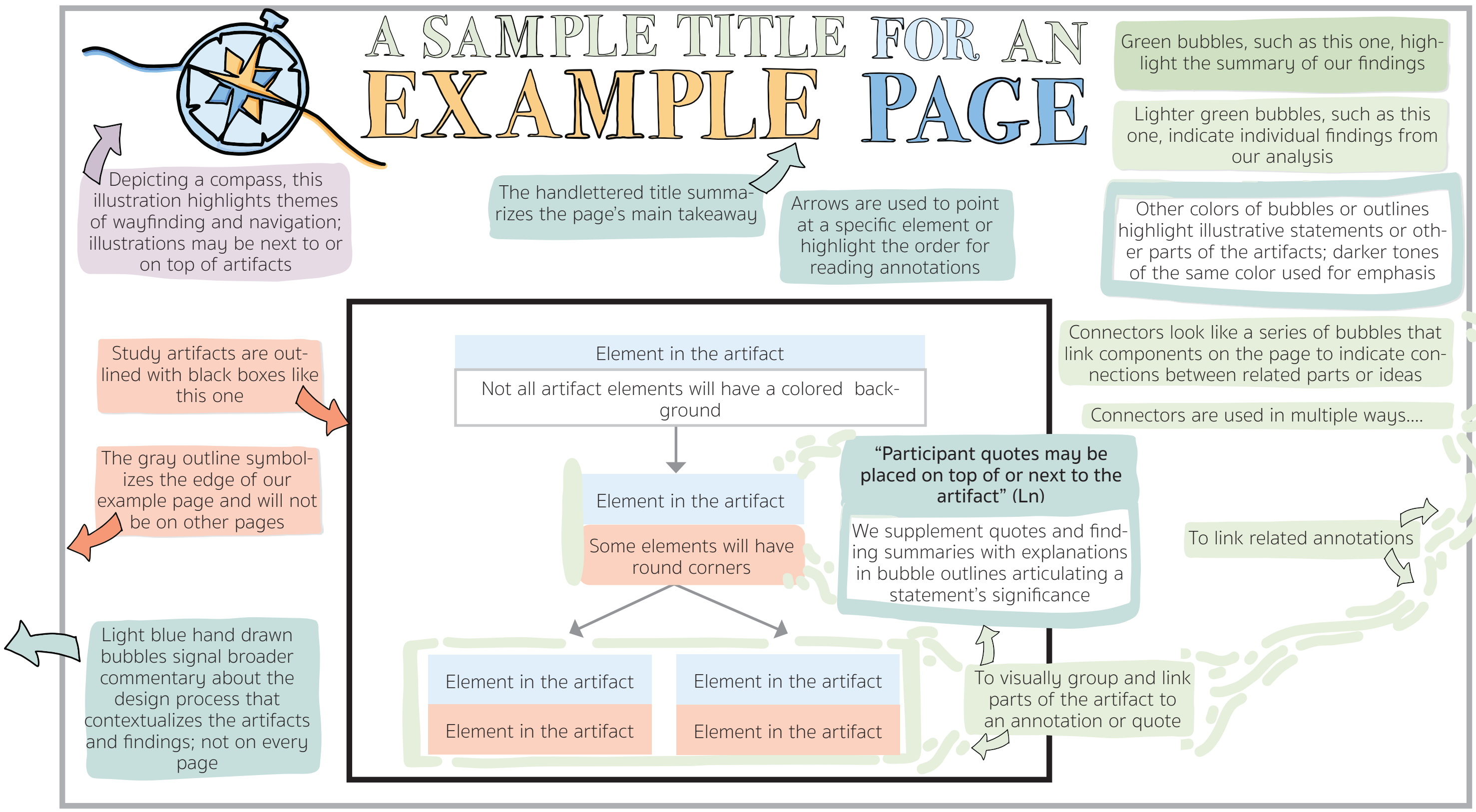

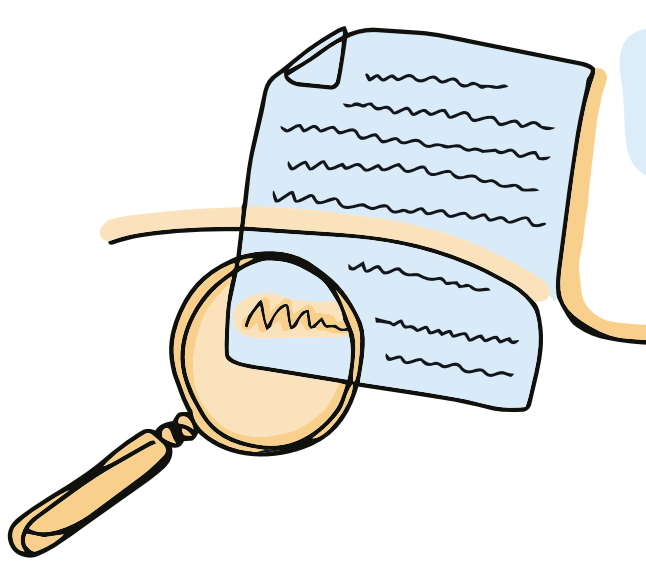

With our project partners, we co-created two visualizations to represent participants' legal analysis. The first visualization (shown on this page) illustrates that legally-relevant information depends on how laws and regulations recognize harm. The second visualization (shown on the following page) illustrates how legal risks depend on dynamics between parties. Later, we discuss how we leveraged the findings from these visualizations to co-create strategies for predicting and managing legal risks.

# FINDING I: LEGALLY RELEVANT INFORMATION DEPENDS ON HOW LAWS AND REGULATIONS RECOGNIZE HARM

Participants said **legally-relevant information depends on how current laws and regulations recognize harm**.

Participants' risk analysis centered on identifying how the situation in the fact pattern could connect to harms recognized by current laws and regulations. Participants noted the risk exposure of different parties is determined by the legal requirements to successfully prove a harm had occurred. Participants described the information in the fact pattern that could satisfy these legal requirements and the additional information they would need to make a determination of the most likely legal risks.

We created a flowchart with the information participants needed to determine legal risk. On this page, we annotate a subsection of this flowchart for the doctors and hospital, highlighting three observations about the information determining legal risks. The same observations hold for other parties, who we discuss more later.

### Doctor

**Are you required to use the medical AI translation tool by your workplace (e.g. hospital) or are you choosing to use it?**

Determines the level of due diligence that's expected

- Yes, I am required to use the tool
  - Less due diligence required from the doctor because the workplace requires the tool
- No, I am not required to use the tool
  - More due diligence required from the doctor to understand how the tool works and prevent harms to patients

**Are you recording someone's voice or image as input into the system?**

Some state privacy laws may require explicit disclosure when live recording audio, taking images, and/or footage for input into a system

**Do you need additional authorization or consent from the patient to use the tool?**

The doctor could face an informed consent claim by the patient for the doctor's failing to adequately disclose the risks about the tool and obtain consent

The doctor could face a privacy violation for using an AI tool that fails to keep patient data confidential in the right context

- If a patient signed a disclosure agreement with the workplace, then the doctor may not (technically) need an additional authorization

  Even if not required, seeking additional consent can help maintain patients' trust with doctors, making a lawsuit less likely
- Patients may be more likely to sue if they feel they are injured from a tool the doctor chose to use on their own, so doctors must take extra care when obtaining consent by explaining potential harm, ensuring patients understand potential negative outcomes, and making sure they sign an authorization document

### Information from multiple people and organizations determines each party's legal risk

Participants needed information from doctors and hospitals to determine if a legally recognized harm could have occurred based on legal standards

We bold the **questions** they would ask each party to elicit legally-relevant information, identifying the underlying legal reason and negative outcomes

### Information can be relevant to the risk of multiple people and organizations

Participants identified where doctors and organizations needed the same or similar information to determine their legal risk.

We drew **connecting arrows** across shared information needs; for readability, we show a subset of the arrows.

For example, the doctor **and** hospital need information about authorized uses of patient data, patient consent, and potential harms

### Legal risk relies on interdependent pieces of information

Participants identified nested questions, meaning the answer to one question indicated subsequent questions they needed to ask.

To reflect this, the flowchart uses **diverging paths of questions**, which we show for the doctor.

### Hospital

⋮

**Do you already have a Business Associates Agreement (BAA) with the vendor (i.e. devel-**

BAA's are contracts used to identify uses of PHI and obligate vendors to comply with regulations; Hospitals need to make vendors sign BAA's to ensure vendors safeguard PHI

**Are the following aspects of how the tool uses PHI covered in the BAA agreement with vendor and subcontractors?**

The High Tech Act means any entity to which the EHR company relayed PHI (and knew it was PHI) would need to comply with HIPAA

If a BAA is not signed or does not include the these required aspects, the hospital is legally and financially responsible for improper disclosures of patient data

**How is patient data and PHI used? Who can acquire patient data and how? What are the potential risks to patients if there is a data breach?**

The HIPAA Security Rule requires a covered entity to do an accurate and thorough risk analysis of the potential threats to electronic health information and how the covered entity will mitigate those threats

If this risk analysis is not correctly completed and data is improperly disclosed, the hospital is required to comply with the Data Breach Notification, resulting in significant costs and potential reputational damages to the hospital

# FINDING 2: LEGAL RISKS DEPEND ON DYNAMICS BETWEEN PARTIES

Participants said legal risks— including the kinds of claims raised, if claims are successful, and what each party may be able to do to change these risks—depended on the decisions and actions of different parties. To identify potential legal risks, lawyers traced how parties connected, noting that while parties further from the patient would be difficult to successfully sue they could still make decisions that deeply impacted patients. Below, we illustrate how legal risks from the translation tool are affected by the distance and dynamics between different parties.

Starting with the central relationship between the doctor and patient, shown in the circle, participants worked backwards to identify the links between stakeholders in the 'ecosystem' of a case

"Look at the possible defendants in a case like an ecosystem..." (L5)

Participants identified hospitals and developers as 'far away' decision makers whose decisions could impact the patient and doctor

"Decisions are made so far away from that central relationship, even though they deeply will affect individual patients" (L1).

Developers are placed furthest away because they do not have direct relationships with patients, making it hardest for patients to bring claims against them

"The further remote an entity is from the individual who is claiming an injury, the harder it is often to bring a successful claim" (L1)

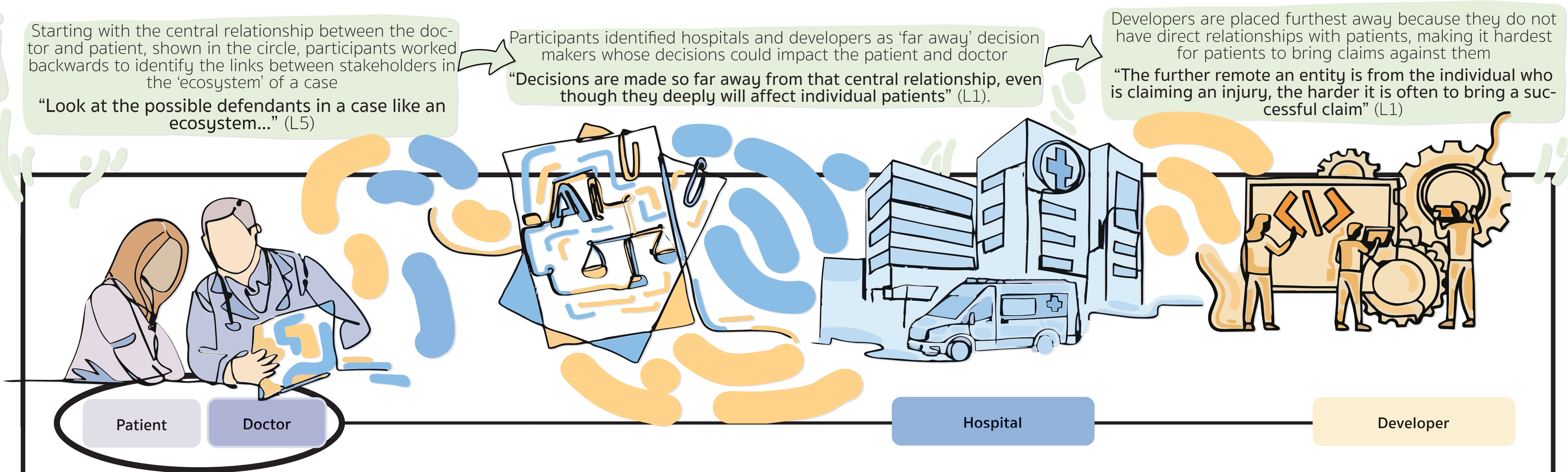

Patients have limited options for avoiding harm and seeking recourse.

"Plaintiffs would have great difficulty bringing any of these claims due to the high injury demonstration requirement" (L1)

Lawyers said patients are most likely to sue doctors when they perceive they are harmed because they are interacting with the doctors face-to-face. However, claims made by patients are not likely to be successful because it would be hard to meet the legal conditions to prove harm by doctors or others. As a result, once the tool is used, patients have limited options for avoiding harm in the first place or achieving recourse via a successful lawsuit.

Doctors face the highest legal risks, and rely on hospitals' work to manage risks.

"But for the most part, there's not a lot they [doctors] can do. You know, they're really relying upon...the diligence and the analysis...[of] the hospital" (L8)

Lawyers said patients are most likely to sue doctors because they interact with them face-to-face. However, doctors rely on the hospital for procedures guiding adoption and use of the tool. Even if claims aren't successful, doctors may experience reputational and psychological

Doctors are incentivized to avoid using the tool if they feel it will cause harms to patients.

**The hospital has the most power to mitigate legal risks of the translation tool**

The hospital's decisions and actions impact developers, doctors, and, ultimately, patients, including mitigating risks. We visually connect hospitals with doctors and developers to show how hospitals are in a position to mitigate risks.

Developers' incentives are to sell their technology to hospitals, and not necessarily aligned with doctors' or patients' interests.

**The hospital has economic power over developers to influence the tool's design, impacting legal risks**

Developers are incentivized to sell their product, so they may be willing to meet performance requirements specified by the hospital

**The hospital defines the policies impacting doctors' legal risks**

Hospitals determine major aspects of how doctors use medical tools, such as approved uses, error reporting procedures, required training, and consent procedures that impact patients' health or perceptions of providers.

**Hospitals face legal, reputational and financial incentives that align their interests with doctors' interests**

"If something goes wrong, a patient's going to sue everybody. They're going to sue the hospital, they're going to sue the doctor..." (L4)

When doctors are sued, hospitals are also typically sued, resulting in financial and reputational consequences. If doctors are sued more often, malpractice insurance premiums may also increase, further raising costs. Finally, hospitals may waste money if they procure and trial a technology only to have their doctors avoid using it.

# FINDING 3: LEGALLY RELEVANT INFORMATION ABOUT AI TOOLS INCLUDE ALGORITHMIC & SOCIAL ELEMENTS

We used our two visualizations—of the legally-relevant information and the dynamics between parties—to co-create strategies for people and organizations to predict and manage legal risks.

On this page, we use the doctor side of the flowchart from Finding 1 to demonstrate how we identified connections to AI tools' design. For readability, we show a selection of the annotations we made of the elements of AI tools' design.

Working with our project partners, we used the information categories about AI tools provided by [15] to annotate our visualization of the legally-relevant information. We find legally-relevant information is related to *algorithmic* **and** *social* elements of AI tools' design.

## Doctor

Are you required to use the medical AI translation tool by your workplace (e.g. hospital) or are you choosing to use it?

Determines the level of due diligence that's expected

Yes, I am required to use the tool

Less due diligence required from the doctor because the workplace requires the tool

No, I am not required to use the tool

More due diligence required from the doctor to understand how the tool works and prevent harms to patients

Are you recording someone's voice or image as input into the system?

Some state privacy laws may require explicit disclosure when live recording audio, taking images, and/or footage for input into a system

Consequences for Stakeholders

Consequences for Users

Data Sourcing and Use

Do you need additional authorization or consent from the patient to use the tool?

The doctor could face an informed consent claim by the patient for the doctor's failing to adequately disclose the risks about the tool and obtain consent

If a patient signed a disclosure agreement with the workplace, then the doctor may not (technically) need an additional authorization

Even if not required, seeking additional consent can help maintain patients' trust with doctors, making a lawsuit less likely

Patients may be more likely to sue if they feel they are injured from a tool the doctor chose to use on their own, so doctors must take extra care when obtaining consent by explaining potential harm, ensuring patients understand potential negative outcomes, and making sure they sign an authorization document

Cross References

Data Sourcing and Use

Consequences for Stakeholders

Model Specifications

Consequences for Users

User Variables

Yellow bubbles show Information Categories from [15] about social processes impacting how AI tools work

Dark blue bubbles show Information Categories from [15] about algorithmic information impacting how AI tools work

The doctor needs algorithmic **and** social information **together** to understand how to use the tool in light of potential legal risks

Doctors need social information about the tool, such as potential data privacy harms for patients (Consequences for Stakeholders) and their legal obligations around communicating harms to patients (Consequences for Users)...

... and algorithmic information about how new patient data is incorporated into the tool (Data Sourcing and Use)...

To understand if they need an additional consent or authorization, doctors need to connect algorithmic elements of AI tools' design with social elements by....

- referencing prior authorizations or disclosure agreements (Cross References) to check if/how they are allowed to input patient data through the tool (Data Sourcing and Use)
- identifying potential impacts on patient health (Consequences for Stakeholders) based on the model's performance as compared to human translators (Model Specifications)
- understanding how to communicate with patients about privacy harms (Consequences for Users) based on how the patient's data connects to the model (Data Sourcing and Use)

On subsequent pages, Findings 4-6 present our visualizations of the strategies that we co-created to understand how to predict and manage risks.

# FINDING 4: PEOPLE AND ORGANIZATIONS MUST PREDICT AND MANAGE RISK THROUGHOUT THE MEDICAL AI TOOL'S CREATION AND USE

With our partners, we analyzed the visualizations based on the participant interviews and co-created strategies for predicting and managing risks. The strategies reflect a key insight from this process: risks do not emerge at a single point but accumulate across the AI tool's adoption. Consequently, we created a new visualization (seen right) of a timeline to reflect how the strategies address risks throughout the four stages of the AI tool's adoption. Because participants indicated hospitals had the greatest ability to impact legal risks, we framed the strategies from their perspective. Later, we discuss how others can reference the same strategies.

### A strategy could contribute to predicting and managing risks through just one stage....

For example, *before procurement* a hospital should double check there is a HIPAA-compliant risk assessment that includes data security risks for the LLM

### ...or across multiple stages

For example, a hospital should require error metrics corresponding to concrete harms....

- *Before procurement* to identify if and how the tool is causing concrete harms, and to avoid procurement if necessary
- *During procurement* to ensure the tool will continue to include and document these measures as a contractual condition
- *While trialing the tool* to ensure the metrics accurately reflect harms that doctors are seeing. Metrics should be updated to ensure new concrete harms are also measured and reflected

### Similar to legal risks, strategies connected to both the algorithmic and social processes impacting AI tools throughout their creation and use

Engaging with physician leadership and including doctors from procurement to deployment in the creation and use of the tool is critical to predicting and mitigating risks. Likewise, collaborating with developers to develop appropriate metrics before procurement can help with predicting and managing risks.

Strategies are shown in purple

We place check marks under the stage(s) in which each strategy should take place

| Strategy | Before Procurement | During Procurement | During Trialing | Deployment & Monitoring |
|---|---|---|---|---|
| Check if the tool is really useful in this context (e.g. Is it better to use a human translator? How do we decide when to use the AI tool?) | ✓ | ✓ | | |
| Check if the tool won't actually create more work or stress for providers, patients, and the hospital | ✓ | ✓ | | |
| Ensure the tool's technology and use by doctors are covered by malpractice insurance | ✓ | ✓ | ✓ | ✓ |
| Engage with physician leadership (e.g. Chief of Division) to foster physician support and engagement for the tool | ✓ | ✓ | ✓ | ✓ |
| Double check the Risk Assessment for the HIPAA Security Rule covers the business associate (i.e. EHR company and third parties), and includes data security risks for the LLM | ✓ | | | |
| Ask the vendor for the Risk Assessment document and make sure it discloses/discusses the risks around the tool | ✓ | ✓ | | |
| Ensure system performance measures/reports include, regularly validate, and document HIPAA compliance measures | ✓ | ✓ | ✓ | ✓ |
| Contractually require and include error metrics to correspond to characteristics of concrete harms as legally defined | ✓ | ✓ | ✓ | ✓ |
| Demonstrate how accuracy metrics were created and measured, including how it reflects potential impacts to patient health (e.g. measure translation errors with different severities; weight aggregate accuracy measures according to potential impacts to patient health) | ✓ | ✓ | ✓ | |
| Demonstrate how accuracy metrics were created and align with measures of quality preceding the tool | ✓ | ✓ | | |
| Collaborate with doctors on reporting measures and processes to more easily identify the presence of new errors | ✓ | ✓ | ✓ | ✓ |
| Ensure the interface features correspond to error reporting protocols and help doctors easily report potential errors | ✓ | ✓ | ✓ | |
| Ensure there are clear protocols to help doctors know when, where, and how to use or report the system | | ✓ | ✓ | ✓ |
| Ensure physician training covers consent procedures, use of the tool error reporting, and other AI governance policies | | | ✓ | ✓ |
| In physician training, error reporting, and all official protocols, clearly communicate the division of responsibility among doctors, the hospital, and developer to protect patient privacy | | | ✓ | ✓ |
| Iteratively pilot language with physicians for use with patients to maintain trust while obtaining consent to use the tool | | | ✓ | ✓ |

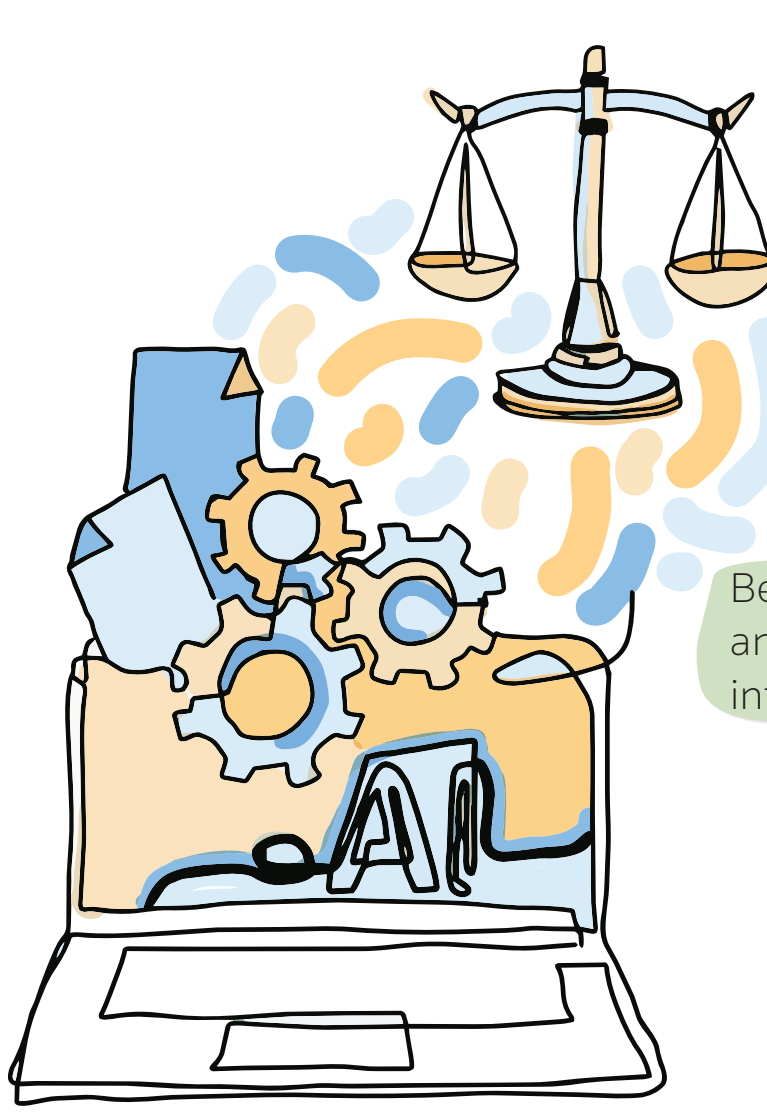

# FINDING 5: PEOPLE AND ORGANIZATIONS MUST PREDICT AND MANAGE RISKS BY DOCUMENTING LEGALLY RELEVANT INFORMATION

Because legally-relevant information is tied to the ways that laws and regulations recognize harm, people and organizations can document this information to predict and manage risks. Some strategies focused on documenting legally-relevant information through legal documents, while others focused on documenting legally-relevant information through the AI tool's algorithmic implementation and processes impacting operation. Below we present a selection of these strategies.

**Critically, the legal documents and tool's design work in tandem**

Together, the legal documents and the tool's design protect the hospital by creating the infrastructures needed to document legally-relevant information *before* harms occur.

A legal document can contractually specify a performance standard corresponding to concrete harms, which could help a hospital predict risks by identifying when legally reocgnized harms are occurring. In turn, the tool's design would correspond to this legal standard, helping the hospital manage related legal risks by actively identifying when and where harms are occurring.

To the right, we use orange and blue annotations to show how the strategies contain a mix of approaches for documenting legally-relevant information, and demonstrate how these approaches should be taken together.

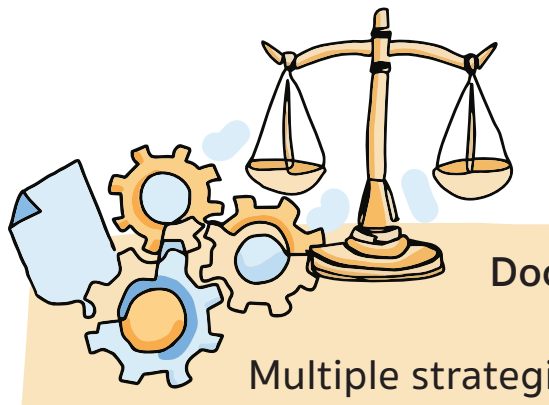

**Documenting through Legal Documents**

**Multiple strategies use legal documents to capture information that can predict or manage risks**

A legal document can record information directly tied to legal requirements

A legal document can contractually specify and obligate developers to meet performance standards related to legal requirements

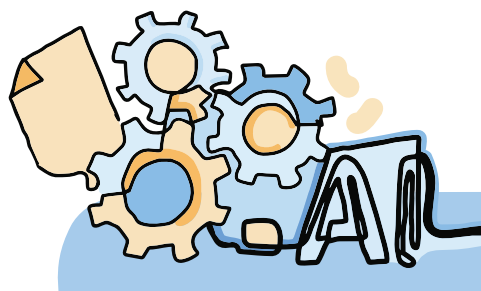

**Documenting through Technology Designs**

**The tool's design needs to record legally-relevant information that directly corresponds to legal documents or requirements**

The tool's error or performance metrics can record information related to legal requirements

The tool could also record information that could be used for legal action, although it doesn't correspond to a legal definition

---

Ask the vendor for the Risk Assessment document and make sure it discloses/discusses the risks around the tool

Double check there is a Risk Assessment for HIPAA Security Rule covers the business associate (i.e. EHR company and third parties), and the HIPAA Risk Assessment covers data security risks related to the LLM

For example, for regulations such as HIPAA, a risk assessment document records legally relevant information corresponding to HIPAA Security Rule about how the tool processes, uses, and shares patient data

Contractually require and include error metrics to correspond to characteristics of concrete harms as legally defined

For example, the hospital can contractually require the vendor to have certain kinds of error metrics targeted at legal harms

As a result, the tool (by design) would record metrics around accuracy corresponding to the legal definition for concrete harms

Ensure system performance measures/reports include, regularly validate, and document HIPAA compliance measures

Similarly, the tool can regularly record and document performance metrics explicitly targeted at HIPAA compliance

Demonstrate how accuracy metrics were created and measured, including how it reflects potential impacts to patient health (e.g. measure translation errors with different severities; weight aggregate accuracy measures according to potential impacts to patient health)

Demonstrate how accuracy metrics were created and align with measures of quality preceding the tool

Measureable aspects of patient health and translation quality don't necessarily correspond to legal definitions, but can be used to meet legal requirements for proving harm in a lawsuit

# FINDING 6: PEOPLE AND ORGANIZATIONS ARE BEST POSITIONED TO PREDICT AND MANAGE RISKS THROUGH INFORMATION ABOUT KNOWN KNOWNS AND KNOWN UNKNOWNS

People and organizations are best positioned to predict and manage risks through information related to harms that are "known knowns" and "known unknowns". By understanding how strategies document different kinds of harms, people and organizations can how they are navigating the underlying landscape of legal risks. Below we use orange and blue annotations to show the strategies address risks from harms that are known knowns and known unknowns.

Double check the Risk Assessment for the HIPAA Security Rule covers the business associate (i.e. EHR company and third parties), and includes data security risks for the LLM

Ask the vendor for the Risk Assessment document and make sure it discloses/discusses the risks around the tool

Ensure system performance measures/reports include, regularly validate, and document HIPAA compliance measures

HIPAA is a federal law that recognizes patient privacy harms. HIPAA applies to medical AI tools in the same way as other medical technology. A hospital can anticipate the need to gather information to predict and manage risk of this harm.

Ensure the tool's technology and use by doctors are covered by malpractice insurance

Malpractice insurance currently protects doctors from financial loss in lawsuits for negligence or injury when treating patients. Medical AI tools may need to be added to insurance, but the strategy is functionally known.

Ensure physician training covers consent procedures, use of the tool error reporting, and other AI governance policies

Similarly, training is a key best practice when trialing and deploying a technology to ensure doctors understand its proper uses and procedures. A hospital can anticipate and incorporate information for consent procedures and error reporting which are especially important for medical AI tools

Demonstrate how accuracy metrics were created and measured, including how it reflects potential impacts to patient health (e.g. measure translation errors with different severities; weight aggregate accuracy measures according to potential impacts to patient health)

Based on current legal standards, a measurable impact on patient health could constitute the basis for a claim. By explicitly connecting metrics to impacts on patient health, a hospital can gather information to predict and manage risk from these kinds of harms before they happen.

Some harms are ***known knowns***: those related to the design of medical AI tools that are more easily anticipated based on current laws and regulations

Strategies gather information in multiple ways in anticipation of these known knowns

Some harms are known knowns based on existing laws and regulations that will apply to medical AI tools

The strategies gather information related to these known regulations

Other harms are known knowns because the same harms can still occur with medical AI tools

These strategies gather information to design processes with current best practices

Finally harms are known knowns because of common ways of establishing a claim

The strategies gather information that can be extrapolated based on prior approaches to establishing a claim

Other harms are ***known unknowns:*** reasonable expectations about where problems could occur in the design of medical AI tools, though the exact nature of the problem is unknown

Strategies gather information in multiple ways to prevent harms that are known unknowns

Some harms are known unknowns because of reasonable expectations about variability in AI tools' performance

The strategies gather information that can serve as early 'warning signs' about previously unencountered errors

Some harms are known unknowns because of common problems with medical technology *development* and *adoption*

The strategies gather information to methodically test and seek out these problems....

...and create a responsive environment to the concerns of those impacted by the technologies, such as doctors and patients

Collaborate with doctors on reporting measures and processes to more easily identify the presence of new errors

Ensure the interface features correspond to error reporting protocols and help doctors easily report potential errors

Ensure there are clear protocols to help doctors know when/where/how to use or report the system in the case that they are made known of an error

AI tools are known to adapt and vary their performance to tailor their outputs to different users or environments. Consequently, a hospital can anticipate errors to occur after the medical AI tool is deployed, even though the exact nature of those errors isn't known. A hospital can anticipate the need to gather information about these errors by coordinating with doctors, and ensuring clear interface features and protocols for error reporting.

Check if the tool is really useful in this context (e.g. Is it better to use a human translator? How do we decide when to use the AI tool?)

Check if the tool won't actually create more work or stress for providers, patients, and the hospital

Many medical technologies are developed that don't add value to clinical settings, which has time pressures and constraints. A hospital can anticipate the need to understand if the medical AI tool is clinically useful throughout its procurement, trialing, and deployment.

Engage with physician leadership (e.g. Chief of Division) to foster physician support and engagement for the tool

Iteratively pilot language with physicians for use with patients to maintain trust while obtaining consent to use the tool

Even when a technology is deployed, many doctors may not use it unless required. A hospital can anticipate the need to gather information about their providers' concerns and support their engagement with the medical AI tool in order to avoid wasted resources.

**PREDICTING & MANAGING LEGAL RISKS**

**Our assets-based, co-design approach surfaced lawyers' tacit knowledge**, enabling us to engage in cross-disciplinary communication to work **alongside** them to predict and manage legal risks. We worked with our two project partners to co-design the interview protocol and fact patterns, analyze the interview data, identify legally-relevant information about medical AI tools, and develop strategies to predict and mitigate risks. Using our interview protocol, we elicited 8 lawyers' skills in risk analysis, identifying the kinds of information needed to analyze legal risks and the dynamics between parties shaping those risks.

Our findings confirm prior work [9, 10, 19] on using **visualizations** in cross-disciplinary design with lawyers, **extending** these findings **to predicting and managing legal risks from AI tools**. Our work demonstrates how visualizations overcome multiple asymmetries in expertise. Creating the flowchart from the interviews helped non-lawyers on our team understand lawyers' tacit knowledge about the structure of the legal risks, such as how risks rely on interdependent pieces of information from multiple people or organizations. The flowchart provided space for our project partners to communicate their own analysis of the interviews, including points of significance. On the flowcharts, information about AI tools' design helped our project partners understand legally-relevant information from the interviews connected to the algorithmic and social processes impacting the design of AI tools.

**For the first time,** to the best of our knowledge**, we leverage fact patterns and visualizations to generatively co-create strategies for predicting and managing legal risks.** Based on our analysis of the dynamics between parties, we identified the hospital as the entity with the most power to mitigate risks from the translation tool. We highlighted how doctors and patients rely on the hospital's work to manage risks. With our project partners, we framed strategies from the perspective of the hospital to reflect their power to impact risks for other stakeholders. Our strategies highlight several takeaways around predicting and managing risks that go beyond the hospital, including how risk mitigation takes place throughout the medical AI tool's creation and use by documenting legally-relevant information. We also highlight how people and organizations are best positioned to address the "known knowns" and "known unknowns" driving legal risks.

This pictorial serves as an example of how **visualizations support communication beyond the co-design team**. By layering the study visualizations and the hand drawn illustrations, our team synthesizes takeaways from the co-design process itself. Through our illustration of the dynamics between parties, we highlight our reflections on the distance between those impacted by legal risks—patients and doctors—and those who are able to make design decisions about medical AI tools—hospitals and developers. Our illustrations amplify our reflections on how legal documents and the AI tool's design document legally-relevant information together. **By drawing out the legal risks**, we can communicate both the **complexity of addressing legal risks** and the **broader implications** of our findings about who is at risk and how those risks can be mitigated.

**TRANSLATING THE STRATEGIES FOR OTHERS**

While this pictorial presented strategies from the hospital's perspective, **other parties** can use them to **understand how** their interests align and how they can collaborate **to lower their legal risks**. For example, doctors would see that they should check that the use of the AI translation tool is covered by their malpractice insurance. Doctors can also talk to their Chief of Division about the tool and ways for their division to influence its procurement and deployment. They can also ask for access to legal counsel or specific trainings around legal risks.

Developers could understand how to better tailor and market to the interests of hospitals and doctors. Administrators for health organizations already identify the need to minimize their legal risks for AI-based technologies [1, 2]. Further, if doctors' legal concerns around medical AI tools are not addressed, they may opt out of using the tools altogether, resulting in wasted resources and lack of benefit from their deployment [5]. Developers can build their products in line with risk mitigation strategies that minimize hospitals' and doctors' risks. For example, developers could see the importance of preemptively conducting risk assessments and performance and metrics tied to HIPAA compliance.

**UNEVENLY DISTRIBUTED RISKS**

Lawyers' strategies for risk mitigation highlighted how **legal risks, and the power to mitigate risks, are not evenly distributed across parties**. While doctors were most likely to be sued, they are not the party with the most power in the procurement, trialing, and development of a medical AI translation tool. Meanwhile, developers play a critical role in how an AI tool is designed, but they were not the party most at risk of legal action. Their distance from patients made it hard to legally demonstrate injury and bring a claim. Hospitals, due to their economic power and ability to make influential decisions around how doctors use the translation AI tool, are the party with the greatest ability to mitigate risks.

Significantly, lawyers' analysis highlighted the **limited options for patients to seek recourse** or mitigate risks from harms due to the translation tool. Patients could (and probably would) raise lawsuits in response to perceived injuries, but they were unlikely to have successful claims or to receive monetary compensation for harms. The main incentives for hospitals to undertake risk mitigation strategies were to avoid harms to doctors (not patients), such as increased legal action against them, and to prevent the waste of resources towards a tool doctors would not use because of unknown legal risks. Consequently, lawyers' primary recommendations for patients was rejecting the use of the AI translation tool.

Even when patients have legal protections, our project partners repeatedly emphasized that patients are **only protected against harms** that are **recognized by laws and regulations**. For example, some harms to patient privacy may not be reflected in HIPAA. Further, even if a hospital follows strategies to protect patient privacy according to HIPAA, patients may still experience other privacy harms. This speaks to existing concerns in the legal community about how some AI tools may cause harms in ways that are not legally recognized, leaving patients without a legal path to recourse [20, 26]. Consequently, **strategies** can predict and manage risks that are legally-recognized, but **do not necessarily provide sufficient or complete protection for patients**.

## REFLECTION AND FUTURE WORK

We see our work as furthering understanding of how to design AI tools in response to legal risks. Using fact patterns, we were able to engage lawyers' expertise around risk analysis to identify legally-relevant information including algorithmic and social processes impacting the design of AI tools. In our example, doctors and hospitals often shared incentives for predicting and managing risks. But we acknowledge the dynamics between parties may vary significantly in other scenarios and require different strategies for predicting and managing risks. For example, more powerful parties may choose to selectively follow strategies in ways that leave more vulnerable parties exposed. Future work should explore how risks and dynamics between parties shift in other domains and identify legal or technological gaps that leave parties unprotected.

Finally, we extend prior work [9, 10] that uses visualizations to support interdisciplinary conversations with lawyers. Future work should explore how visualizations can help other diverse stakeholders communicate with lawyers to identify concrete recommendations for AI tools and shape risk mitigation in light of legal infrastructures.

## ACKNOWLEDGEMENTS

MANY THANKS TO BENJAMIN SUNDHOLM FOR SHARING HIS EXPERTISE AND FOR HIS SHARP AND THOUGHTFUL FEEDBACK ABOUT THE METHODOLOGY AND CONTENTS OF THIS WORK. WE ALSO GRATEFULLY ACKNOWLEDGE OUR PARTICIPANTS FOR THEIR TIME, ATTENTION, AND CARE.